\documentclass{emulateapj}
\usepackage{apjfonts}

\slugcomment{To appear in ApJ September 20, 2008}

\newlength{\figsize}
\newcommand{\Figure}[2]
           {\centering\leavevmode\includegraphics[width=#2,clip]{#1.ps}}
\newcommand{\eq}[1]{\begin{equation} #1 \end{equation}}

\newcommand\D     {{\sf DUSTY}}
\newcommand\about {\hbox{$\sim$}}
\newcommand{\E}[1]{\hbox{$10^{#1}$}}
\newcommand\x     {\hbox{$\times$}}
\newcommand\mic   {\hbox{$\mu$m}}

\newcommand\alb   {\hbox{$\varpi_\lambda$}}
\newcommand\Ac    {\hbox{$A_{\rm c}$}}
\newcommand\Bl    {\hbox{$B_{\lambda}$}}
\newcommand\ec    {\hbox{$\eta_{\rm C}$}}
\newcommand\fel   {\hbox{$f_{\rm e\lambda}$}}
\newcommand\F     {\hbox{$F_{\lambda}$}}
\newcommand\Fe    {\hbox{$F_{\rm e}$}}

\newcommand\Jl    {\hbox{$J_{\lambda}$}}
\newcommand\I     {\hbox{$I_{\lambda}$}}
\newcommand\IC    {\hbox{$I^{\rm C}_{\lambda}$}}

\newcommand\mH    {\hbox{$m_{\rm H}$}}
\newcommand\nc    {\hbox{$n_{\rm C}$}}
\newcommand\Nc    {\hbox{$N_{\rm C}$}}
\newcommand\N     {\hbox{$\cal N$}}
\newcommand\NT    {\hbox{${\cal N}_{\rm T}$}}
\newcommand\No    {\hbox{${\cal N}_0$}}
\newcommand\Pesc  {\hbox{$P_{\rm esc}$}}
\newcommand\pAGN  {\hbox{$p_{\rm AGN}$}}
\newcommand\qal   {\hbox{$q_{\rm a\lambda}$}}
\newcommand\qae   {\hbox{$q_{\rm ae}$}}
\newcommand\qP    {\hbox{$q_{\rm aP}$}}

\newcommand\Rc    {\hbox{$R_{\rm c}$}}
\newcommand\Rd    {\hbox{$R_{\rm d}$}}
\newcommand\Sl    {\hbox{$S_{\rm c,\lambda}$}}
\newcommand\Sd    {\hbox{$S^{\rm d}_{\rm c,\lambda}$}}
\newcommand\Si    {\hbox{$S^{\rm i}_{\rm c,\lambda}$}}
\newcommand\thin  {\hbox{$\theta_{\rm i}$}}
\newcommand\thout {\hbox{$\theta_{\rm o}$}}
\newcommand\Tm    {\hbox{$T_{\rm max}$}}
\newcommand\tl    {\hbox{$\tau_{\rm \lambda}$}}
\newcommand\tV    {\hbox{$\tau_{\rm V}$}}
\newcommand\Vc    {\hbox{$V_{\rm c}$}}
\newcommand\lh    {\hbox{$\lambda_{\rm h}$}}
\newcommand\lu    {\hbox{$\lambda_{\rm u}$}}
\newcommand\lRJ   {\hbox{$\lambda_{\rm RJ}$}}

 \shorttitle{AGN Dusty Tori: I. Handling of Clumpy Media}
 \shortauthors{Nenkova et al}

\begin{document}

\title{AGN Dusty Tori: I. Handling of Clumpy Media}

\author{Maia Nenkova\altaffilmark{1},
        Matthew M. Sirocky\altaffilmark{2},
        \v{Z}eljko Ivezi\'c\altaffilmark{3}
        and Moshe Elitzur\altaffilmark{2}}


\altaffiltext{1}{Seneca College, Toronto,
                 ON M2J 2X5, Canada; maia.nenkova@senecac.on.ca}
\altaffiltext{2}{Department of Physics and Astronomy, University of Kentucky,
                Lexington, KY 40506-0055; sirockmm@pa.uky.edu,
                moshe@pa.uky.edu}
\altaffiltext{3}{Department of Astronomy, University of Washington,
                 Seattle, WA 98105; ivezic@astro.washington.edu}

\begin{abstract}
According to unified schemes of Active Galactic Nuclei (AGN), the central
engine is surrounded by dusty, optically thick clouds in a toroidal structure.
We have recently developed a formalism that for the first time takes proper
account of the clumpy nature of the AGN torus. We now provide a detailed report
of our findings in a two-paper series. Here we present our general formalism
for radiative transfer in clumpy media and construct its building blocks for
the AGN problem --- the source functions of individual dusty clouds heated by
the AGN radiation field. We show that a fundamental difference from smooth
density distributions is that in a clumpy medium, a large range of dust
temperatures coexist at the same distance from the radiation central source.
This distinct property explains the low dust temperatures found close to the
nucleus of NGC1068 in 10 \mic\ interferometric observations. We find that
irrespective of the overall geometry, a clumpy dust distribution shows only
moderate variation in its spectral energy distribution, and the 10\mic\
absorption feature is never deep. Furthermore, the X-ray attenuating column
density is widely scattered around the column density that characterizes the IR
emission. All of these properties are characteristic of AGN observations. The
assembly of clouds into AGN tori and comparison with observations is presented
in the companion paper.

\end{abstract}


\section{INTRODUCTION}

Although there are numerous AGN classes, a unified scheme has been emerging
steadily \citep[e.g.,][]{Ski93, Ski02, Urry95}. The nuclear activity is powered
by a supermassive black hole and its accretion disk, and this central engine is
surrounded by a dusty toroidal structure. Much of the observed diversity is
simply explained as the result of viewing this axisymmetric geometry from
different angles. The torus provides anisotropic obscuration of the central
region so that sources viewed face-on are recognized as ``type 1'', those
observed edge-on are ``type 2''. From basic considerations, \cite{Krolik88}
conclude that the torus is likely to consist of a large number of individually
very optically thick dusty clouds. Indeed, \cite{Tristram07} find that their
VLTI interferometic observations of the Circinus AGN ``provide strong evidence
for a clumpy or filamentary dust structure''. A fundamental difference between
clumpy and smooth density distributions is that radiation can propagate freely
between different regions of an optically thick medium when it is clumpy, but
not otherwise. However, because of the difficulties in handling clumpy media,
models of the torus IR emission, beginning with \cite{PK92}, utilized smooth
density distributions instead. \cite{RR95} noted the importance of
incorporating clumpiness for realistic modeling, but did not offer a formalism
for handling clumpy media.

We have recently developed such a formalism and presented initial reports of
our modeling of AGN clumpy tori in \cite*{NIE02}, \cite*{Elitzur04}, and
\cite{Elitzur06, Elitzur07}. We now offer a detailed exposition of our
formalism and its application to AGN observations. Because of the wealth of
details, the presentation is broken into two parts. In this paper we describe
the clumpiness formalism and construct the source functions for the emission
from individual optically thick dusty clouds, the building blocks of the AGN
torus. In a companion paper \cite[part II hereafter]{AGN2} we present the
assembly of these building blocks into a torus and the application to AGN
observations.

\section{CLUMPY MEDIA}

We start by presenting a general formalism for handling clumpy media. For
completeness, the full formalism is described here, including results that are
utilized only in part II.

\begin{figure}[ht]
 \Figure{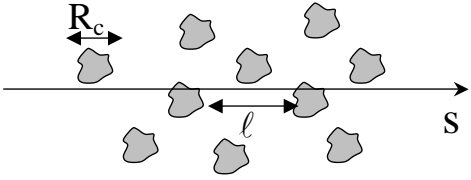}{0.8\figsize}

\figcaption{A region populated by clouds of size $\Rc$. Along ray $s$, the
photon mean-free-path is $\ell$ and the number of clouds per unit length is
\Nc\ = $\ell^{-1}$. \label{ClumpyMed}
}
\end{figure}

Consider a region where the matter is concentrated in clouds (see Fig.\
\ref{ClumpyMed}). For simplicity we take all clouds to be identical;
generalizing our results to a mixture of cloud properties is straightforward
\citep*{Conway05}. Individual clouds are characterized by their size \Rc, the
cloud distribution is specified by the number of clouds per unit volume \nc.
Denote by \Vc\ the volume of a single cloud and by $\phi$ the volume filling
factor of all clouds, i.e., the fraction of the overall volume that they
occupy. We consider the medium to be clumpy whenever
\eq{\label{eq:clumpy}
 \phi = \nc\Vc  \ll 1\,.
}
In contrast, the matter distribution is smooth, or continuous, when $\phi
\simeq 1$. It is useful to introduce the number of clouds per unit length
\eq{\label{eq:mfp}
    \Nc = \nc\Ac = \ell^{-1}
}
where \Ac\ is the cloud cross-sectional area and $\ell$ is the photon mean free
path for travel between clouds (Fig.\ \ref{ClumpyMed}). Since $\Vc \simeq
\Ac\Rc$, the clumpiness condition (eq.\ \ref{eq:clumpy}) is equivalent to
\eq{\label{eq:clumpy2}
    \phi = \Nc\Rc \ll 1, \qquad \hbox{or}\quad \Rc \ll \ell.
}
The clumpiness criterion is met when the mean free path between clouds greatly
exceeds the cloud size.



\subsection{Emission from a clumpy medium}
\label{sec:formalism}

When the clumpiness criterion is met, each cloud can be considered a
``mega-particle'', a point source of intensity \Sl\ and optical depth
$\tau_\lambda$. The intensity at an arbitrary point $s$ along a given path can
then be calculated by applying the formal solution of radiative transfer to the
clumpy medium. The intensity generated in segment $ds'$ around a previous point
$s'$ along the path is $\Sl(s')\Nc(s')ds'$. Denote by $\N(s',s) =
\int_{s'}^s\Nc ds$ the mean number of clouds between $s'$ and $s$ and by
$\Pesc(s',s)$ the probability that the radiation from $s'$ will reach $s$
without absorption by intervening clouds. \cite{Natta84} show that when the
number of clouds is distributed around the mean $\N(s',s)$ according to Poisson
statistics,
\eq{\label{eq:Pesc}
  \Pesc(s',s) = e^{-t_\lambda(s',s)},            \qquad \hbox{where}\quad
  t_\lambda(s',s)  = \N(s',s)( 1 - e^{-\tau_\lambda} );
}
the Appendix discusses the validity of the Poisson distribution and presents
the derivation of this result. Its intuitive meaning is straightforward in two
limiting cases. When $\tau_\lambda < 1$ we have $t_\lambda(s',s) \simeq
\N(s',s)\tau_\lambda$, the overall optical depth between points $s'$ and $s$;
that is, clumpiness is irrelevant when individual clouds are optically thin,
and the region can be handled with the smooth-density approach. It is important
to note that $\N\tau_\lambda$ can be large---the only requirement for this
limit is that each cloud be optically thin. The opposite limit $\tau_\lambda >
1$ gives $\Pesc(s',s) \simeq e^{-{\cal N}(s',s)}$. Even though each cloud is
optically thick, a photon can still travel between two points  along the path
if it avoids all the clouds in between.

With this result, the intensity at $s$ generated by clouds along the given ray
is
\eq{\label{eq:IC}
  \IC(s) = \int^s e^{-t_\lambda(s',s)} \Sl(s')\Nc(s') ds'
}
This relation is the exact analog of the formal solution of standard radiative
transfer in continuous media, to which it reverts when the cloud sizes are
decreased to the point that they become microscopic particles; in that case
$\tau \ll 1$ for each particle and $\ell^{-1}$ is the standard absorption
coefficient. The only difference between the clumpy and continuous cases is
that optical depth is replaced by its effective equivalent $t_\lambda$ and
absorption coefficient is replaced by the number of clouds per unit length \Nc.
It should be noted that, because of the particulate composition of matter, the
solution of radiative transfer is always valid only in a statistical sense,
corresponding in principle to the intensity averaged along the same path over
an ensemble of many sources with identical average properties. However, in the
smooth-density case the large number of dust particles along any path (a column
with unity optical depth contains \about\ \E{10} particles when made up of 0.1
\mic\ dust grains) implies such small relative fluctuations around the mean
that the statistical nature can be ignored and the results considered
deterministic. For example, the intensity contours produced by a smooth-density
spherical distribution are perfectly circular for all practical purposes. In
the clumpy case, on the other hand, the small number of particles along the
path can lead to substantial deviations from the mean intensity \IC. In
contrast with the smooth density case, the intensity of a spherical cloud
distribution can fluctuate significantly around concentric circles even though
the distribution average properties are constant on such circles. The flux
emitted from the enclosed circular area involves areal integration which
smoothes out and reduces these fluctuations, so deviations of individual
spectral energy distributions (SED) from their average are expected to be
smaller than the brightness fluctuations. In general, the SED can be expected
to have lower fluctuations than the brightness in clumpy distributions with a
smoothly behaved function \Nc. Estimating fluctuations is beyond the scope of
our formalism, which is predicated on averages at the outset. An assessment of
fluctuations would require an extended theoretical apparatus or analysis of
Monte-Carlo simulations, such as those described by \cite{Hoenig06}.

After characterizing the clumpy medium by the cloud size \Rc\ and the volume
density of clouds \nc\ we introduced an equivalent set of two other independent
variables---the volume filling factor $\phi$ and the number of clouds per unit
length \Nc. Our final expression for the intensity does not involve $\phi$;
only \Nc\ enters. A complete formalism that would not invoke the assumption
$\phi \ll 1$ would lead to a series expansion in powers of $\phi$, and the
expression we derived in eq.\ \ref{eq:IC} is the zeroth order term in that
expansion. In fact, detailed Monte Carlo simulations show that, to within a few
percents, this expression describes adequately clumpy media with $\phi$ as
large as 0.1 (Conway, Elitzur \& Parra, in preparation). Since the intensity
calculations are independent of the volume filling factor, their results do not
provide any information on this quantity, nor do they provide separate
information on either \Rc\ or \nc, only on \Nc. In complete analogy, the
radiative transfer problem for smooth density distributions does not involve
separately the size of the dust grains or their volume density, only the
combination $n_{\rm d}\sigma_{\rm d}$, which determines the absorption
coefficient and which is equivalent to \Nc.

Equation \ref{eq:IC} accounts only for the radiation generated along the path
by the clouds themselves. A path containing a background source, such as the
line of sight to the AGN, requires different handling since averaging is then
meaningless. For such a line of sight, the intensity $I_k$ generated when there
are exactly $k$ intervening clouds ($k = 0, 1, 2, \dots$) has a Poisson
probability $P_k$. The only meaningful quantities that can be deduced from
modeling in this case are tabulations of the intensities $I_k$ and their
associated probabilities $P_k$; that is, the probability is $P_0$ that the
intensity is $I_0$ (the unobscured AGN), $P_1$ that it is $I_1$, etc. An actual
source will correspond to one particular member of this probability
distribution.

\subsection{The cloud distribution}
The only distribution required for intensity calculation is \Nc, the number of
clouds per unit length. Our interest is primarily in cloud distributions with
axial symmetry in which \Nc\ depends only on the distance $r$ from the
distribution center and the angle $\beta$ from the equatorial plane (the
complementary of the standard polar angle). The total number of clouds along
radial rays slanted at angle $\beta$ is, on average, $\NT(\beta) =
\int\!\Nc(r,\beta)\,dr$. It is convenient to introduce as a free parameter the
mean of the total number of clouds along radial rays in the equatorial plane,
\No\ = \NT(0). The angular profile $\NT(\beta)/\No$ is expected to decline as
$\beta$ increases away from the equatorial plane in AGN tori.

\subsection{Total mass in clouds}
The mass of a single cloud can be written as $\simeq \mH N_{\rm c,H}\Ac$, where
$N_{\rm c,H}$ denotes column density and \mH\ is the hydrogen nucleus mass.
With the aid of eq.\ \ref{eq:mfp} and introducing the distribution profile
$\ec(r,\beta) = (1/\No)\Nc(r,\beta)$\footnote{The profile \ec\ obeys the
normalization $\int\!\ec(r,0)\,dr = 1$; note that \ec\ and \Nc\ have dimensions
of inverse length while \No\ is dimensionless.}, the total mass in clouds is
\eq{\label{eq:M_C}
    M_{\rm C} = \mH N_{\rm c,H}\No\int\!\!\ec\,dV
}
where the integration is over the entire volume populated by clouds. Note that
$N_{\rm c,H}\No$ is the mean overall column density in the equatorial plane.
The total mass of a smooth distribution with the same equatorial column density
would be given by the same expression except that the profile \ec\ would then
describe the normalized gas density distribution.

The result in equation \ref{eq:M_C} for the overall mass in the clouds is
independent of the filling factor $\phi$. The only property of an individual
cloud that enters into this expression is its column density $N_{\rm c,H}$,
which is directly related to its optical depth. {\em The cloud size \Rc\ is
irrelevant so long as it meets the clumpiness criterion eq.\ \ref{eq:clumpy2}.}

\subsection{Total number of clouds}
Similar to the intensity, the overall mass involves the number of clouds per
unit length \Nc, not the cloud volume density \nc. The latter quantity enters
in the calculation of the total number of clouds $n_{\rm tot} = \int\nc dV$.
Expressing $n_{\rm tot}$ too in terms of \Nc\ brings in the cloud size because
$n_{\rm tot} = \int(\Nc/\Ac)dV$. With $\Ac \simeq R_{\rm c}^2$ and $\Rc =
\phi/\Nc$,
\eq{
    n_{\rm tot} = \N_0^3\int{\eta_{\rm c}^3\over\phi^2}\, dV.
}
Among the quantities of interest, $n_{\rm tot}$ is the only one dependent upon
the volume filling factor. Self-consistency of our formalism requires $\phi \ll
1$ to ensure the clumpiness criterion and $n_{\rm tot} \gg \No$ to validate the
Poisson distribution along paths through the source (see Appendix). Since \ec\
obeys the normalization $\int\!\ec(r,0)\,dr = 1$, $n_{\rm tot}$ is of order
\about\ $\N_0^3/\phi^2$ and the two requirements are mutually consistent.

\subsection{Covering Factors}
Various AGN studies have invoked the concept of a ``covering factor'' in
different contexts. The covering factor is generally understood as the fraction
of the sky at the AGN center covered by obscuring material. This is the same as
the fraction of randomly distributed observers whose view to the center is
blocked. Therefore the average covering factor of a random sample of AGN is the
same as the fraction, $f_2$, of type 2 sources in this sample; that is, $f_2$
is the average of the fraction of the AGN radiation absorbed by obscuring
clouds. Denote the spectral shape of the AGN radiation by \fel, normalized
according to $\int\fel d\lambda = 1$. The fraction of the AGN luminosity
escaping through a spherical shell of radius $r$ centered on the nucleus is, on
average,
\eq{\label{eq:pAGN}
    \pAGN(r) = \int_0^1\!\! d\sin\beta \int\! d\lambda\,
               \fel\, P_{\lambda,\rm esc}(r,\beta)\,,
}
where $P_{\lambda,\rm esc}(r,\beta)$ is the probability for a photon of
wavelength $\lambda$ emitted by the AGN in direction $\beta$ to reach radius
$r$ (eq.\ \ref{eq:Pesc}). Therefore $f_2 = 1 - \pAGN(R_{\rm out})$, where
$R_{\rm out}$ is the torus outer radius, a relation that holds independent of
the magnitude of the optical depth of individual clouds. The spectral
integration involves all wavelengths in general, and is limited to the relevant
spectral range in cases of specific obscuring bands.  When the clouds are
optically thick to the bulk of the AGN radiation, $P_{\lambda,\rm esc}(R_{\rm
out},\beta) \simeq e^{-{\cal N}_{\rm T}(\beta)}$, independent of $\lambda$
(eq.\ \ref{eq:Pesc}). Utilizing the normalization of \fel\ then yields
\eq{\label{eq:f2}
    f_2 = 1 - \int_0^{\pi/2}e^{-{\cal N_{\rm T}}(\beta)} \cos\beta d\beta.
}
In a recent study \cite{Maiolino07a} defined the ``dust covering factor'' as
the ratio of thermal infrared emission to the primary AGN radiation. Since the
bulk of the AGN output is in the optical/UV, the radiation absorbed by the
torus clouds must be re-radiated in infrared, therefore this dust covering
factor is the fraction $f_2$ for optical/UV. In X-rays, the fraction $f_2$ is
usually derived from the statistics of sources that have at least one obscuring
cloud, whether Compton-thick or not, along the line of sight to the AGN. Since
the probability to miss all clouds is $e^{-{\cal N}_{\rm T}}$, equation
\ref{eq:f2} holds in this case, too, with \NT\ the overall number of X-ray
obscuring clouds, whether dusty or dust-free. Therefore, the
\citeauthor{Maiolino07a}\ covering factors for dust and X-rays properly
correspond to the $f_2$ fractions for optical/UV and X-rays, respectively, and
can be used for a meaningful comparison between similar quantities. Although
the actual comparison is subject to many observational uncertainties that
affect differently the two wavelength regimes, its formal, fundamental premise
is valid.

Other definitions of a ``covering factor'' do not always correspond to the
fraction of obscured sources, and sometimes even lead to values that exceed
unity. \cite{Krolik88} introduce a ``covering fraction'' $C$ such that the
average column density near the equatorial plane is $C\,N_{\rm c,H}$.
Therefore, their covering factor is \No, the average number of clouds along
radial equatorial rays. Broad and narrow line emission from AGN is frequently
calculated from an expression similar to eq.\ \ref{eq:IC} without considering
cloud obscuration, i.e., in the limit $t_\lambda = 0$
\citep[e.g.,][]{Netzer90}. The cloud distribution is taken as spherical and the
normalization of the total line flux is obtained from a ``covering factor''.
Similar to \citeauthor{Krolik88}, this covering factor is \NT, the total number
of clouds along radial rays \citep[see eq.\ 8 in][]{Kaspi99}. Since a spherical
distribution has $f_2 = 1 - e^{-{\cal N}_{\rm T}}$ (eq.\ \ref{eq:f2}), this
covering factor coincides with $f_2$ only when $\NT \ll 1$. Note that a
spherical distribution with \NT\ = 1 would have a unity covering factor
according to this definition, because a cloud is encountered in every direction
on average, when in fact $f_2$ is only 0.63 in this case.

The emission line covering factor is the cloud number \NT, not the fraction
$f_2$. This covering factor is obtained from the analysis of type 1 spectra
under very specific approximations. The population of clouds dominating the
calculation under these assumptions can be quite different from that
controlling the AGN obscuration, and this covering factor can differ a lot from
the fraction $f_2$ obtained from source statistics. As this discussion shows,
the concept of a covering factor has not been well defined in the literature,
referring to different quantities in different contexts.  When this concept is
invoked, a proper definition requires calculation of probability from the
detailed cloud distribution function, similar to the one for $f_2$ above (eq.
\ref{eq:f2}).


\begin{figure}[ht]
 \Figure{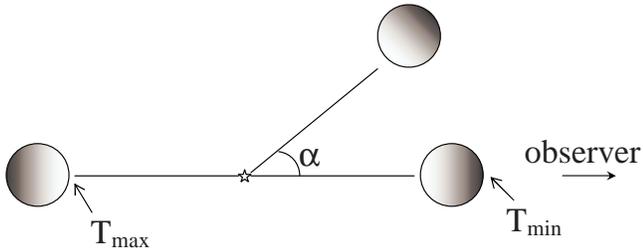}{\figsize}

\caption{Clouds at a fixed distance from the AGN at various position angles
$\alpha$ with respect to the observer direction. The dust temperature varies on
the surface of an optically thick cloud from $T_{\rm max}$ on the illuminated
face to $T_{\rm min}$ on the dark side. Therefore the emission is anisotropic
and $\alpha$ determines the visible fraction of the illuminated surface area.}
 \label{clumps}
\end{figure}


\section{SOURCE FUNCTION}

The formalism just described is general.\footnote{For its application to line
emission from a clumpy medium see Conway et al 2005.} Its application to a
specific radiative process requires modeling of the source function for a
single cloud. Here we apply this formalism to dusty torus clouds heated by the
central AGN radiation. We divide the clouds into two classes. For large optical
depths, clouds directly facing the AGN will have a higher temperature on the
illuminated face than on other surface areas (Fig.\ \ref{clumps}). Their
emission is therefore strongly anisotropic, and the corresponding source
function \Sd\ depends on both distance $r$ from the AGN and the angle $\alpha$
between the directions of the cloud and observer\footnote{Denote by $i$ the
angle to the observer direction from the axis of symmetry. A cloud at angle
$\beta$ from the equatorial plane and azimuthal angle $\varphi$ from the
axis--observer plane has
  $$  \cos\alpha = \cos\beta\cos\varphi\sin i + \sin\beta\cos i $$
}.
Clouds whose view of the AGN is blocked by another cloud will be heated only
indirectly by all other clouds. Heating of these clouds by diffuse radiation
is, in general, much more even, and the $\alpha$-dependence is expected to be
much weaker for their source function \Si\ than for \Sd. At location
$(r,\beta)$, the mean number of clouds to the AGN is $\N(r,\beta) =
\int_0^r\!\Nc(r,\beta)\,dr$ and the probability for unhindered view of the AGN
is $p(r,\beta) = e^{-{\cal N}(r,\beta)}$.  The general expression for the cloud
overall source function at that location is thus
\eq{\label{eq:S}
      \Sl(r,\alpha,\beta) =
      p(r,\beta)\Sd(r,\alpha) + \left[1 - p(r,\beta)\right]\Si(r,\alpha,\beta)
}
We now describe our detailed calculations of the source functions of the two
classes of clouds.

\subsection{Directly illuminated clouds}

\begin{figure}[ht]
 \Figure{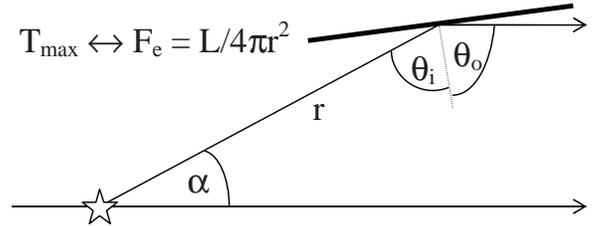}{0.9\figsize}

\caption{A synthetic cloud at distance $r$ from the AGN and position angle
$\alpha$ from the observer direction is constructed by averaging the emission
from a slab with the same optical depth over all slab orientations (see eq.\
\ref{eq:clump}). The local illuminating flux, \Fe, and maximal temperature on
the cloud surface, \Tm, are uniquely related to each other (see
\S\ref{sec:Tdust}).
}
 \label{fig:slab}
\end{figure}

Clouds come in all shapes and forms. But whatever its shape, when the size of a
cloud directly illuminated by the AGN is much smaller than $r$, it is
indistinguishable from a flat patch with the same overall optical depth.
Indeed, all calculations of line emission from AGN always model the line
emitting clouds as slabs illuminated by the central engine along the normal to
their surface. However, a fundamental difference between a flat slab and an
actual cloud of the same optical thickness is that the former presents to an
observer either its bright or dark faces while the latter generally presents a
combination of both. To account for this effect we construct ``synthetic
clouds'' by averaging the emission from an illuminated slab over all possible
slab orientations (figure \ref{fig:slab}). This procedure utilizes an exact
solution of radiative transfer for externally illuminated slabs, and we proceed
now to discuss the properties of an irradiated dusty slab; the synthetic clouds
are constructed from the slab solutions in \S\ref{sec:Synthesis}.


\subsubsection{Slab radiative transfer}
\label{slab}

Thanks to the planar symmetry, the slab radiative transfer equation can be
formulated completely in optical depth space; {\em neither the density profile
nor the slab geometrical thickness are relevant}. In the case of dusty slabs
this invariance extends to the temperature equation because the only
attenuation of the heating radiation comes from radiative interactions.
Therefore the slab radiative transfer problem is fully specified by the optical
depth $\tl = q_\lambda\tV$, where \tV\ is the dust optical depth across the
slab at visual wavelengths and $q_\lambda$ is the relative efficiency factor at
wavelength $\lambda$. The radiative transfer equation along a ray at angle
$\arccos\mu$ to the slab normal is
\eq{
         \mu{d\I \over d\tl} = S_\lambda(\tl) - \I(\tl)
}
where \tl\ is measured in the normal direction from the illuminated face and
$S_\lambda$ is the source function. For dust albedo \alb\ and isotropic
scattering, $S_\lambda = (1-\alb)\Bl( T) + \alb\Jl$ where \Jl\ is the angle
averaged intensity and $T$ the dust temperature, obtained at each point in the
slab from the coupled equation of radiative equilibrium.

We solve the slab radiative transfer problem with the 1D code \D\
\citep*{DUSTY}.  The code takes advantage of the scaling properties of the
radiative transfer problem for dust absorption, emission and scattering
\citep[IE97 hereafter]{IE97}. The solution determines both the radiation field
and the dust temperature profile in the slab. The dust grains are spherical
with size distribution from \cite*{MRN}. The composition has a standard
Galactic mix of 53\% silicates and 47\% graphite. The optical constants for
graphite are from \cite{Draine03}, for silicate from the \cite*{OHM} ``cold
dust'', which produces better agreement with observations of the 10 and 18
\mic\ silicate features \citep{Sirocky08}. From the optical constants \D\
calculates the absorption and scattering cross-sections using the Mie theory
and replaces the grain mixture with a single-type composite grain whose
radiative constants are constructed from the mix average. This method
reproduces adequately full calculations of grain mixtures, especially when
optical depths are large \citep[e.g.,][]{EfRR94, Wolf03}. The handling of the
dust optical properties is exact in this approach, the only approximation is in
replacing the temperatures of the different grain components at each point in
the slab with a single average. \cite{Wolf03} finds that the temperatures of
different grains are within $\about \pm10\%$ of that obtained in the mean grain
approximation, and that these deviations disappear altogether when $\tau > 10$.
Henceforth, the term ``dust temperature'' implies this mean temperature of the
mixture.


\begin{figure}[ht]
 \Figure{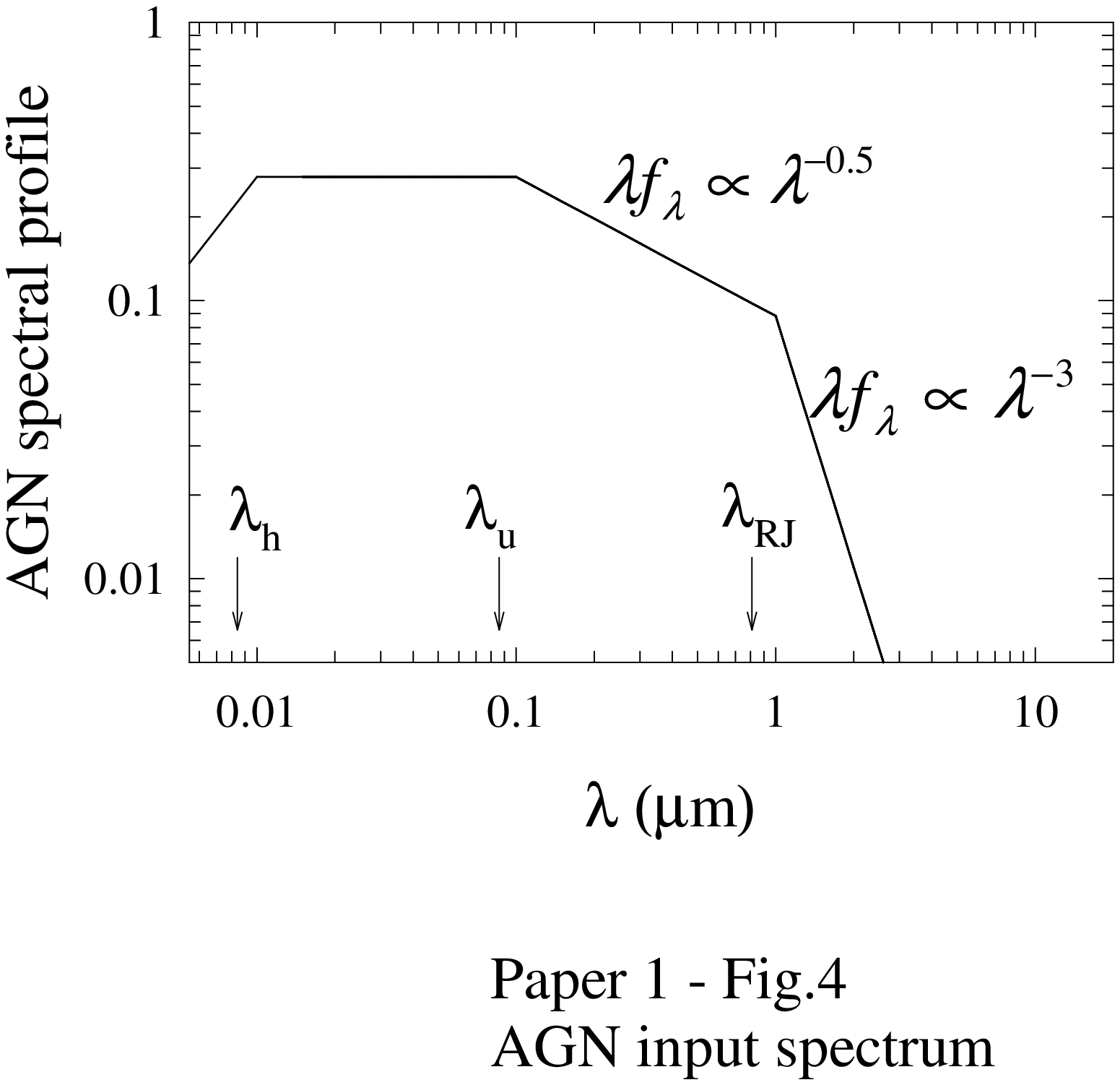}{\figsize}

\caption{The spectral shape of the AGN illuminating radiation for the standard
set of parameters (see eq.\ \ref{eq:fAGN}).} \label{Fig:fAGN}
\end{figure}


The slab is illuminated by the AGN flux $F_{\rm e,\lambda}$ at an angle \thin\
from the normal. For isotropic AGN emission with luminosity $L$, the bolometric
flux at the slab position is
\eq{\label{eq:Fe}
           \Fe = {L \over 4\pi r^2 }
}
The illuminating flux is characterized by \Fe\ and by the AGN normalized SED
$\fel = F_{\rm e,\lambda}/\Fe$. Following \cite{RR95}, we employ the piecewise
power-law distribution\footnote{Note that $F_\nu \propto \nu^{\alpha}$ implies
$\lambda\F \propto \lambda^{-(\alpha + 1)}$}
\eq{\label{eq:fAGN}
 \lambda\fel \propto \cases{
      \lambda^{1.2} & \hskip 0.2in  $         \lambda \le \lh$   \cr
      \rm const     & \hskip 0.2in  $\lh  \le \lambda \le \lu$   \cr
      \lambda^{-p}  & \hskip 0.2in  $\lu  \le \lambda \le \lRJ$  \cr
      \lambda^{-3}  & \hskip 0.2in  $\lRJ \le \lambda$           \cr
      }
}
with the following set of parameters: \lh\ = 0.01\mic, \lu\ = 0.1\mic, \lRJ\ =
1\mic\ and $p$ = 0.5 (see fig.\ \ref{Fig:fAGN}).  The effects of varying the
parameters from this standard set are discussed below in \S\ref{sec:AGN_input}.


\begin{figure}[ht]
 \Figure{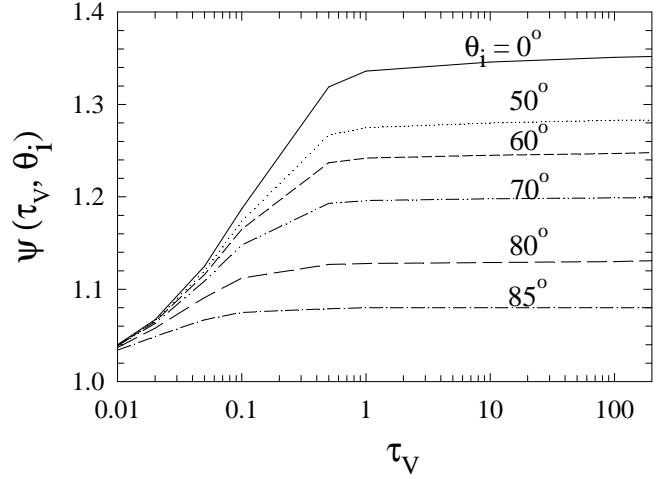}{\figsize}

\caption{The effects of optical thickness \tV\ (normal to the slab sides) and
illumination angle \thin\ on the temperature of the slab illuminated surface.
Shown is the variation of the function $\psi$ (see eq.\ \ref{eq:T-Fe}) with
\tV\ at various \thin, as marked. The dust in all figures has standard Galactic
composition, the temperature is the mixture average. The flatness of $\psi$ for
$\tV \ga 1$ implies that the dust temperature is the same on the illuminated
face of all optically thick slabs.} \label{Fig:Psitau}
\end{figure}


\subsubsection{Dust temperature}
\label{sec:Tdust}

When $\tV \ll 1$ the diffuse radiation inside the slab is negligible. The dust
temperature is the same everywhere in the slab and depends only on \Fe; it does
not depend on either \tV\ or the illumination angle \thin\ (so long as the slab
remains optically thin along the slanted direction). The same holds for any
externally heated cloud as long as it is optically thin and its size is much
smaller that the distance to the source, and the slab geometry faithfully
mimics this aspect. By contrast, when the radiation source is embedded inside a
smooth density distribution (e.g., a spherical shell), the dust temperature
does vary even in the optically thin case because of the spatial dilution of
radiation with distance from the source.

When $\tV > 1$ the slab diffuse radiation contributes significantly to dust
heating. If \qal\ is the efficiency factor for absorption, let $\qae =
\int\!\qal\fel d\lambda$ and \qP\ the corresponding average with the Planck
spectral shape. From radiative equilibrium, the dust temperature at the slab
illuminated boundary obeys (IE97)
\eq{\label{eq:T-Fe}
        \sigma T^4\qP(T) = \frac14\Fe\qae\psi(\tV,\thin)
 }
The function $\psi$ introduced here accounts for the contribution of the
diffuse radiation to the energy density, and can be obtained only from a
detailed solution of the radiative transfer problem; it is normalized according
to $\psi(0,\thin) = 1$ to recover the temperature of optically thin dust
(IE97\footnote{Note that IE97 utilized the function $\Psi = (\qae/\qP)\psi$}).
Irrespective of the actual value of the dust temperature on the illuminated
face, the function $\psi$ fully describes the effects of illumination angle and
slab overall optical depth on that temperature.  Figure \ref{Fig:Psitau} shows
the variation of $\psi$ with both \tV\ and \thin. The dependence on either
variable is weak. Even for normal illumination, where it is the largest, $\psi$
does not exceed 1.35, reaching its maximum at $\tV \ga 1$; i.e., the
contribution of the diffuse component to the surface heating is no more than
35\% of the direct heating by the external source. This behavior is markedly
different from the spherical case where $\psi$ increases monotonically with
\tV\ without bound (IE97). The reason is the fundamental difference in the
relation between radiative flux and energy density in the two geometries.  In a
spherical shell, the energy density can increase indefinitely in the inner
cavity because there is no energy transport there (from symmetry, the diffuse
flux vanishes in the cavity). As the optical depth increases, the radiation
trapped in the cavity dominates the dust heating on the shell inner boundary,
leading to an unbounded increase of $\psi$ with \tV\ (the same surface
temperature requires a smaller \Fe). In contrast, a slab cannot store energy
anywhere. When a slab optical thickness increases, any increase in diffuse
energy density is accompanied by a commensurate flux increase, producing the
saturation effect evident in the figure. As a result, the temperature on the
illuminated face is the same, independent of optical depth when $\tV \ga 1$.


\begin{figure}[ht]
 \Figure{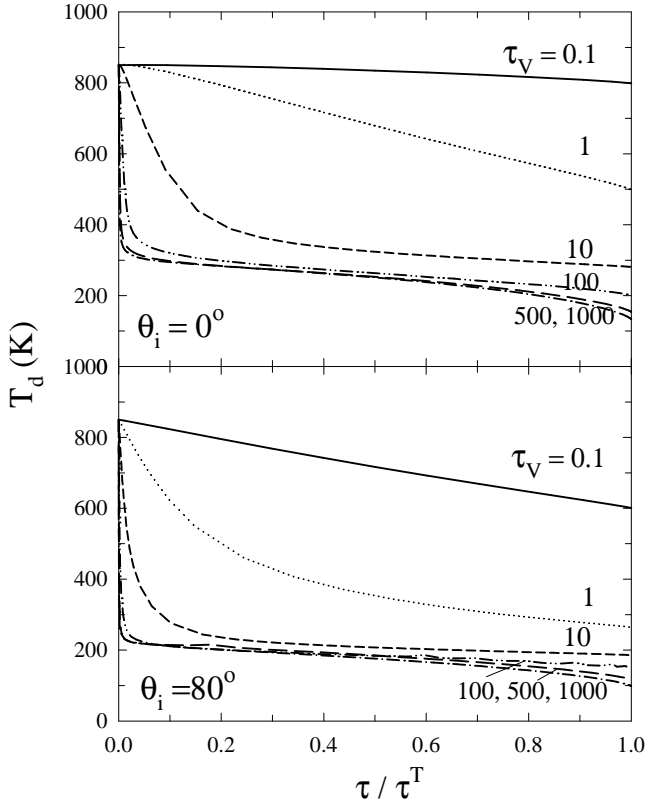}{\figsize}

\caption{Temperature variation inside slabs of various \tV, as marked, for two
illumination angles. The variable $\tau/\tau^{\rm T}$ is thickness into the
slab from the illuminated face, where the temperature is 850 K for all the
displayed slabs.} \label{Fig:Td-tau}
\end{figure}

\begin{figure}[ht]
 \Figure{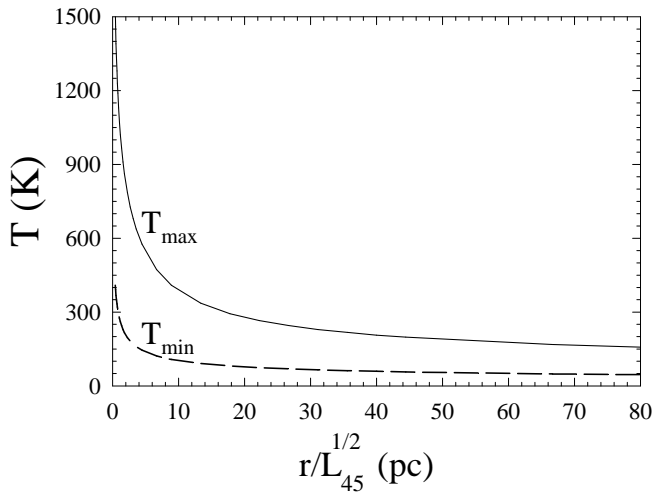}{\figsize}

\caption{An optically thick dusty slab at distance $r$ from an AGN with
luminosity $L_{45} = L$/(\E{45} erg\,s$^{-1}$) is irradiated normal to the
surface. The dust temperature is $T_{\rm max}$ on the slab illuminated face,
$T_{\rm min}$ on the dark side. The curve for $T_{\rm max}$ is applicable
whenever $\tV > 1$, the one for $T_{\rm min}$ when $\tV > 10$.
} \label{Fig:Tm-r}
\end{figure}


Figure \ref{Fig:Td-tau} displays the temperature variation inside
representative slabs of varying optical thickness heated to a surface
temperature of 850 K. After a sharp drop near the illuminated face, the
temperature is nearly constant throughout the rest of the slab for optical
depths exceeding \tV\ \about\ 10. There is little difference between the
profiles at large overall optical depths---all slabs with \tV\ $\ge$ 100
produce nearly indistinguishable profiles. Increasing \thin\ has a similar
effect to increasing \tV\ because of the slanted illumination. The temperature
decline near the illuminated surface reflects the exponential attenuation of
the external heating radiation. Once this radiation is extinguished, the dust
temperature in the rest of the slab is maintained by the diffuse radiation. And
since the bolometric flux is constant and must be radiated from the slab dark
side, this temperature is uniquely determined by the external flux and the
grain optical properties. It is independent of the overall optical depth once
that optical depth exceeds unity for all wavelengths around the Planck peak.
This behavior is in stark contrast with all centrally heated dust distributions
where the flux radiated from the outer boundary decreases as the surface area
increases with overall size, leading to ever declining dust temperatures.

As is evident from figure \ref{Fig:Psitau}, $\psi(\tV,\thin)$ is nearly
constant as a function of \tV\ for $\tV \ge 1$; deviations are within the 5\%
accuracy of our numerical results. Therefore, at a given distance $r$ from the
AGN and a given inclination \thin, all optically thick slabs are heated to the
same surface temperature. And as is evident from figure \ref{Fig:Td-tau}, they
also have essentially the same dust temperature on their dark face when $\tV
> 10$. Figure \ref{Fig:Tm-r} displays the variation of these two surface
temperatures with distance from the AGN for optically thick slabs with normal
illumination. The displayed relations can be described with simple analytic
approximations. The temperature $T_{\rm max}$ on the illuminated face reaches
1500 K at distance $R_{\rm d} \simeq 0.4\,L_{45}^{1/2}$ pc, where $L_{45} =
L$/(\E{45} erg\,s$^{-1}$). Introduce $\rho = r/\Rd$, then
\begin{eqnarray}
\label{eq:r-T}
  \nonumber
  T_{\rm max} &\simeq& \cases{
            1500\cdot (1/\rho)^{0.39}\ \hbox{K} & $\rho \le 9$  \cr \cr
  \phantom{1}640\cdot (9/\rho)^{0.45}\ \hbox{K} & $\rho  >  9$  \cr}  \\
  \nonumber
  \\
  T_{\rm min} &\simeq& \phantom{1}400\cdot (1/\rho)^{0.42}\ \hbox{K}
\end{eqnarray}
These temperatures bracket the range of dust temperatures on the surface of a
cloud at distance $\rho$ from the AGN (see figure \ref{clumps}). Because the
dust temperature is determined by radiative equilibrium, the large disparity
between different surface areas cannot be equalized by cloud rotation. The dust
is primarily heated by absorption of short wavelength radiation in electronic
transitions, with typical time scales of \E{-6} sec, and cools via vibrational
transitions within seconds. Therefore dust radiative equilibrium is established
instantaneously when compared with any dynamical time scale. It is also
important to note that the gas has no effect on the dust temperature as long as
the density is $\la$ \E{12} cm$^{-3}$.

All the clouds illustrated in figure \ref{clumps} are at the same distance from
the AGN, yet their radiation toward the observer involves a mix of a wide range
of surface temperatures. This result has profound implications for the torus
emission. The 10 \mic\ dust emission in NGC1068 has been recently resolved in
VLTI interferometry by \cite{Jaffe04}, who analyzed their observations with a
model containing a compact ($r \la$ 0.5 pc) hot ($>$ 800 K) core and cooler
(320 K) dust extending to $r \simeq$ 1.7 pc. \cite{Poncelet06} reanalyzed the
same data with slightly different assumptions and reached similar conclusions
--- the coolest component in their model has an average temperature of 226 K
and extends to $r$ = 2.7 pc. As noted by the latter authors, the close
proximity of regions with dust temperature of $>$ 800 K on one hand and \about
200--300 K on the other is a most puzzling, fundamental problem. Clumpiness
provides a natural solution, thanks to the spatial collocation of widely
different dust temperatures. With a bolometric luminosity of \about\ 2\x\E{45}
erg s$^{-1}$ for NGC1068 \citep{Mason06}, the dust temperature in an optically
thick cloud at $r$ = 2 pc is 960 K on the bright side but only 247 K on the
dark side, declining further to 209 K at 3 pc. Indeed, the temperatures deduced
in the model synthesis of the VLTI data fall inside the range covered by the
cloud surface temperatures at the derived distances. \cite{Schartmann05} have
recently modeled the NGC1068 torus with multi-grain dust in a smooth density
distribution and found that, although the different dust components span at the
same location a range of temperatures, this range is smaller than required.
They conclude that even with this refinement their model cannot explain the
VLTI interferometric observations and that the clumpy structure of the dust
configuration must be included in realistic modeling. The same effect is found
in the Circinus AGN, where \cite{Tristram07} conclude similarly that the
temperature distribution inferred from their VLTI observations provides strong
evidence for a clumpy or filamentary dust structure.

At a given distance $r$, the highest surface temperature is obtained for slab
orientation normal to the radius vector. This maximal temperature has a
one-to-one correspondence with \Fe. Thanks to this correspondence, \Tm\ can be
used instead of \Fe\ to characterize the boundary condition of the solution.

\subsubsection{Emerging intensity from a slab}

\begin{figure}[ht]
 \Figure{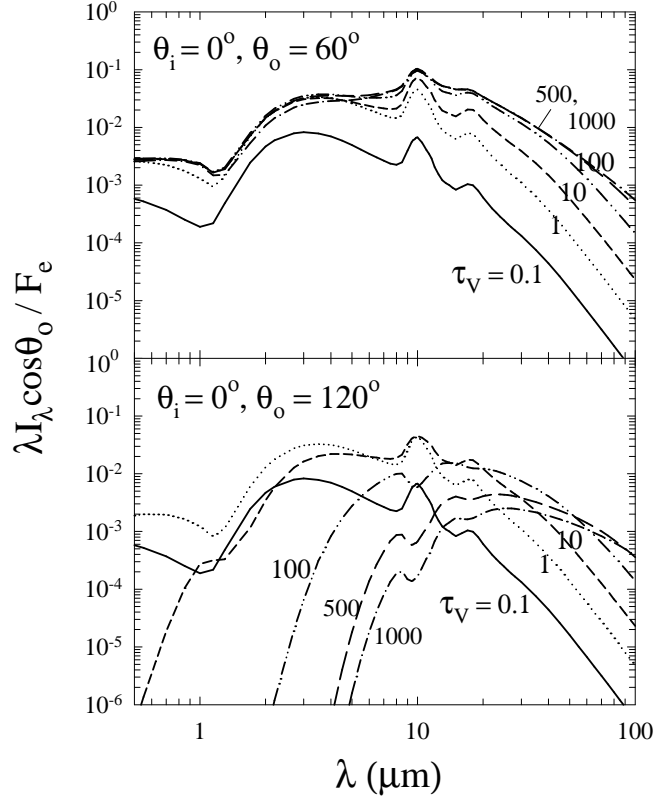}{\figsize}
\caption{Optical depth dependence of SED of slabs illuminated by normal
radiation to \Tm\ = 850 K. In the top and bottom panels the observer direction
is 60$\degr$ from slab normal on the illuminated and dark sides, respectively.}
\label{Fig:int-tauV}
\end{figure}

\begin{figure}[ht]
 \Figure{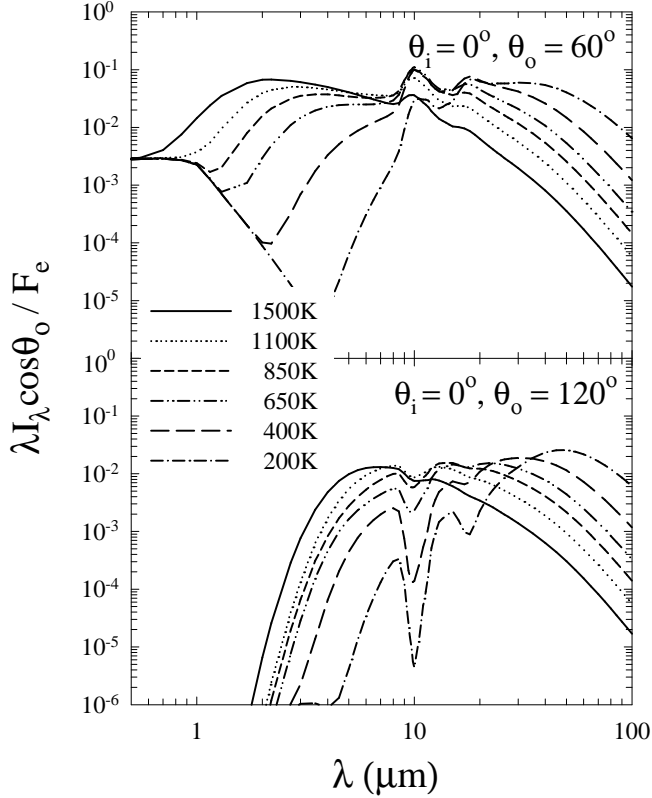}{\figsize}
\caption{The SED of slabs with \tV\ = 100 normally illuminated to surface
temperature \Tm, as marked. All other parameters are the same as in Fig.
\ref{Fig:int-tauV}.} \label{int-T1}
\end{figure}

\begin{figure}[ht]
 \Figure{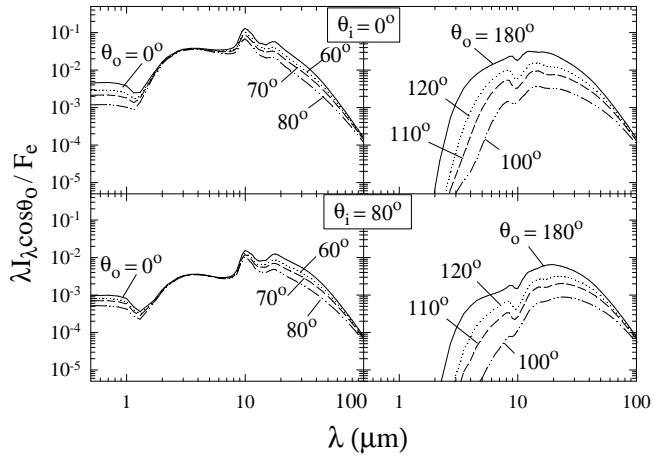}{\figsize}

\caption{Dependence of slab SED on illumination and observation angles \thin\
and \thout. The observer is on the illuminated side in the left panels, on the
dark side in the right panels. All slabs have \tV\ = 100, the external flux
\Fe\ corresponds to \Tm\ = 850 K (the surface temperature that a normally
illuminated slab would have at that location).
} \label{4int-theta}
\end{figure}

The spectral shape of the radiation emerging from the slab depends on \tV, \Tm,
\thin\ and the angle \thout\ between the observer direction and the slab normal
(see figure \ref{fig:slab}). Figures \ref{Fig:int-tauV}--\ref{4int-theta}
display the dependence of the slab SED on these four parameters. Each figure
displays separately the SEDs for the illuminated and dark sides of the slab,
which are fundamentally different from each other. Every SED is scaled with
\Fe, the AGN bolometric flux at the slab location (see eq.\ \ref{eq:Fe}).


Figure \ref{Fig:int-tauV} displays the \tV\ dependence of the SED for
representative parameters.  On the illuminated side the SED is dominated by
scattering and hot emission from the surface layer. The silicate 10\mic\
feature appears always in emission, although its contrast decreases with \tV.
On the dark side, the feature displays the behavior familiar from spherical
models (e.g., IE97): emission at small \tV, switching to absorption at large
\tV\ when the hot radiation from the illuminated region propagates through
optically thick cool layers. However, the absorption feature is never as deep
as in the spherical case , reflecting the flat temperature profile (fig.
\ref{Fig:Td-tau}).


Figure \ref{int-T1} shows the variation of the SED with the temperature of the
illuminated surface. The change in temperature affects the dust emission,
shifting it to longer wavelengths as $T$ decreases. This modifies the 10\mic\
silicate feature. When the emission peak moves past 10\mic, the silicate
feature starts disappearing from the SED. Short wavelengths ($\lambda \la
1\mic$) are dominated by the scattered component. They are visible only on the
bright side and are unaffected by this change.


Figure \ref{4int-theta} shows the effects of the illumination and viewing
angles. Moving the observer's direction away from the slab normal has a similar
effect to increasing the slab optical depth. Varying the illumination angle
affects the attenuation of the external heating radiation.


\begin{figure}[ht]
 \Figure{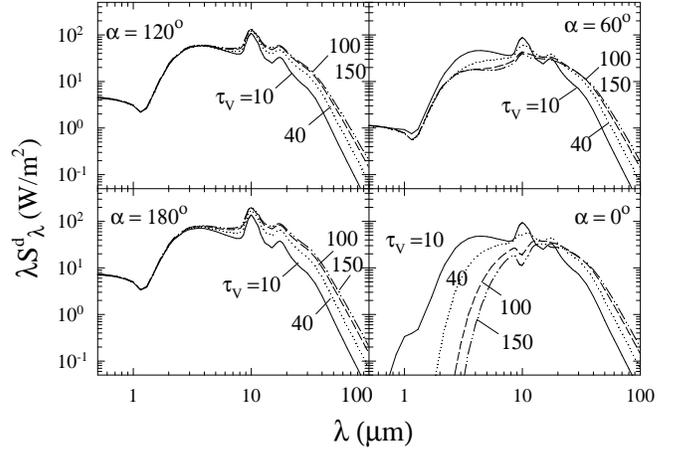}{\figsize}

\caption{Dependence of the source function of directly illuminated clouds on
optical depth. The clouds are located at $r$-equivalent temperature \Tm\ = 850
K and position angles as marked; $\alpha = 0\degr$ is directly in front of the
AGN, $\alpha = 180\degr$ directly behind (see fig.\ \ref{clumps}). }
 \label{Fig:Sdir-tau}
\end{figure}


\subsubsection{Cloud source function \Sd}
\label{sec:Synthesis}

A slab-like patch observed at angle \thout\ from a large distance appears as a
point whose intensity (flux per solid angle) is $\I(\thout)\cos\thout$. At a
distance $r$ from the AGN, corresponding to temperature \Tm, and position angle
$\alpha$ we construct the source function for a ``synthetic cloud'' with
optical depth \tV\ from
\eq{\label{eq:clump}
  \Sd(\Tm,\tV,\alpha) = \langle I(\Tm,\tV,\thin,\thout)\cos\thout\rangle
}
Here $\I(\Tm,\tV,\thin,\thout)$ is the intensity of a slab with the listed
parameters and the brackets denote averaging over all possible slab
orientations (Fig.\ \ref{fig:slab}). The fraction of slabs with observable
bright side is the same as the visible fraction of the illuminated area on the
surface of a spherical-like cloud.

\begin{figure*}[ht]
\begin{minipage}{\textwidth}
 \Figure{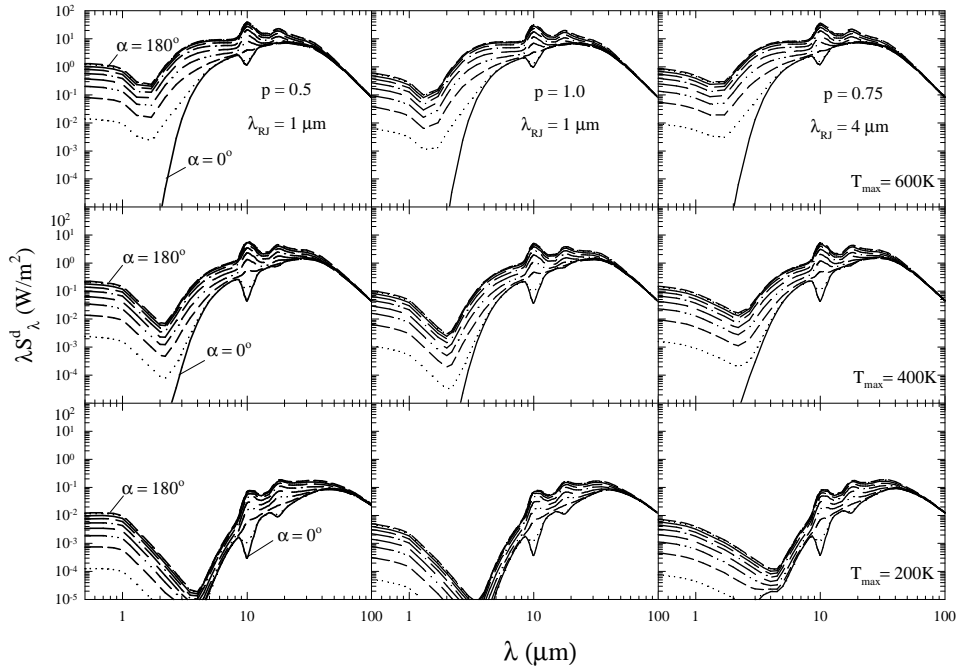}{0.7\hsize}

\caption{Dependence of the source function for directly illuminated clouds with
\tV\ = 100 on position angle $\alpha$, varied at $20\degr$ intervals. Each row
of panels corresponds to $r$-equivalent temperature \Tm\ as marked. Each column
corresponds to a different set of $p$ and \lRJ\ (see eq.\ \ref{eq:fAGN}) for
the illuminating spectral shape, with the central one corresponding to the
standard parameters .}
 \label{Fig:Sdir-all}
\end{minipage}
\end{figure*}

Figure \ref{Fig:Sdir-tau} displays the dependence of the source function \Sd\
on optical depth for clouds located at the same distance at a number of
representative position angles around the AGN. At $\alpha = 0\degr$, only the
dark side of the cloud is visible (see fig.\ \ref{clumps}). Increasing $\alpha$
exposes to the observer a larger fraction of the illuminated area until it is
fully visible at $\alpha = 180\degr$. This explains the emergence of short
wavelengths and the switch from absorption to emission of the 10 \mic\ silicate
feature as $\alpha$ increases. The same behavior is evident also in figure
\ref{Fig:Sdir-all}, which provides a more detailed coverage of the
$\alpha$-dependence of \Sd\ as well as additional temperatures.

At all position angles, increasing the optical depth of a single cloud beyond
$\tV \sim 100$ has only a minimal effect on the spectral shape. And except for
the short wavelengths at $\alpha = 0$, the SED similarity extends down almost
to clouds with \tV\ = 10. This behavior reflects the saturation of the slab
temperature profile (see fig.\ \ref{Fig:Td-tau}) and is a realistic depiction
of the situation inside an actual cloud heated from outside by a distant
source. An important consequence of the flatness of the temperature profile
through most of the cloud is that the silicate absorption feature never becomes
deep --- a uniform temperature source cannot produce an absorption feature at
all. This is fundamentally different from continuous dust distributions
surrounding the heating source, where the temperature keeps decreasing with
radial distance and the absorption feature keeps getting deeper as the overall
optical depth increases until finally the entire wavelength region is
suppressed by self absorption. In an externally heated cloud, on the other
hand, the depth of the absorption feature is set by the contrast between
temperatures on the two faces. When both $T_{\rm max}$ and $T_{\rm min}$ are
higher than \about\ 300 K (the Planck-equivalent of 10\mic, roughly) the entire
slab contributes to 10\mic\ emission; when both are lower, no region emits
appreciably at this wavelength. The absorption feature is deepest when the hot
face is warmer than \about\ 300 K and the cool side is cooler, maximizing the
contrast between the emitting and absorbing layers. The solutions displayed in
figure \ref{Fig:Sdir-all} for \Tm\ = 400K and 200K show the deepest absorption
features produced by directly illuminated clouds for any combination of the
parameters.


\subsubsection{The input spectral shape}
\label{sec:AGN_input}

All previous descriptions of the spectral shape of the AGN input continuum are
summarized by the piecewise power law in eq.\ \ref{eq:fAGN} \citep[see][and
references therein]{RR95}. The continuum shape shortward of 0.1\mic\ (1000\AA)
is uncertain; the slope of the spectral falloff between optical and X-rays has
been determined for many sources but the location \lh\ of the turndown from
flat $\lambda\fel$ toward high frequencies remains unknown. Fortunately, this
spectral region has no effect on the shape of the dust SED. It affects only the
normalization of the relation between luminosity and dust temperature (eq.\
\ref{eq:r-T}), and we find this to be only \about\ 1\% effect when \lh\
increases from our standard 0.01\mic\ all the way to 0.03\mic.

The impact of the optical---near-IR region is much more significant. Since dust
cannot be hotter than its sublimation temperature (\about\ 1500 K), it emits
predominantly at $\lambda \ga$ 2--3\mic. Shorter wavelengths involve the Wien
tail of emission by the hottest dust and scattering of the AGN radiation, and
thus reflect directly the input spectrum. We characterize the AGN emission in
this region by the spectral index $p$ and the wavelength \lRJ, which marks the
onset of the Rayleigh-Jeans tail $\fel \propto \lambda^{-4}$. This wavelength
corresponds roughly to the lowest temperature on the  accretion disk of the
central engine. The spectral index $p$ has been determined for 6868 quasars
studied in the Sloan Digital Sky Survey \citep{Ivezic02}. Its distribution
covers the range $-1 \le p \le 1.5$ and has a flat peak at $0.5 \le p \le 0.8$.
Direct observational determination of \lRJ\ requires near-IR spectral studies
of type 1 AGN with angular resolution better than 0\farcs01 for a source
distance of 10 Mpc. Such observations are yet to be performed. Promising
indirect methods for separating the near-IR emission into its torus and disk
components include measurements of continuum polarization \citep{Kishimoto05}
and multiple regression analysis of time variability \citep{Tomita06}.

Figure \ref{Fig:Sdir-all} shows SEDs for clouds illuminated by radiation with
$p$ = 0.5 and $p$ = 1, two cases that bracket the peak of the observed
$p$-distribution. The input spectrum makes no difference at wavelengths
dominated by dust emission, $\lambda \ga$ 2\mic. Shorter wavelengths are
dominated by the AGN scattered radiation, and the displayed results reflect the
difference in input radiation in this spectral region. Although not large,
these differences are important because IR spectral indices that include at
least one wavelength shorter than 2\mic\ are a common tool of spectral analysis
\citep[e.g.,][]{Almudena03}. The figure also shows an intermediate slope, $p$ =
0.75, with \lRJ\ increased from 1\mic\ to 4\mic. As is evident from the figure,
such an increase will also produce an observable effect on IR spectral indices.


\begin{figure}[ht]
 \Figure{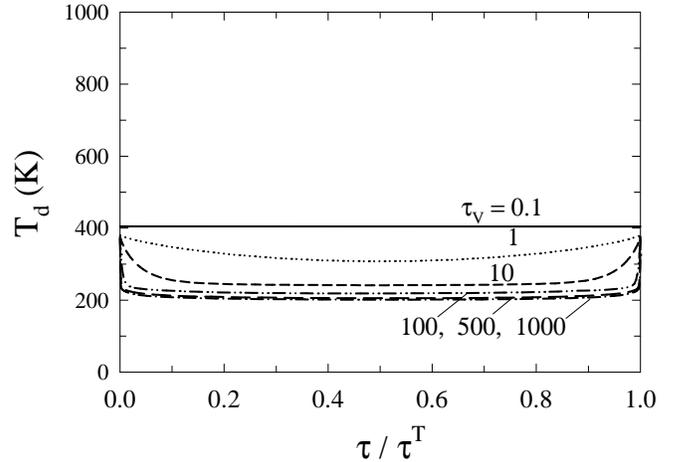}{\figsize}

\caption{Temperature variation inside slabs of various \tV, as marked, heated
indirectly by diffuse radiation at distance from the AGN with equivalent
temperature \Tm\ = 850 K (cf fig.\ \ref{Fig:Td-tau}). The variable
$\tau/\tau^{\rm T}$ is thickness into the slab from one face.
}
 \label{Fig:Tdiff}
\end{figure}



\begin{figure}[ht]
 \Figure{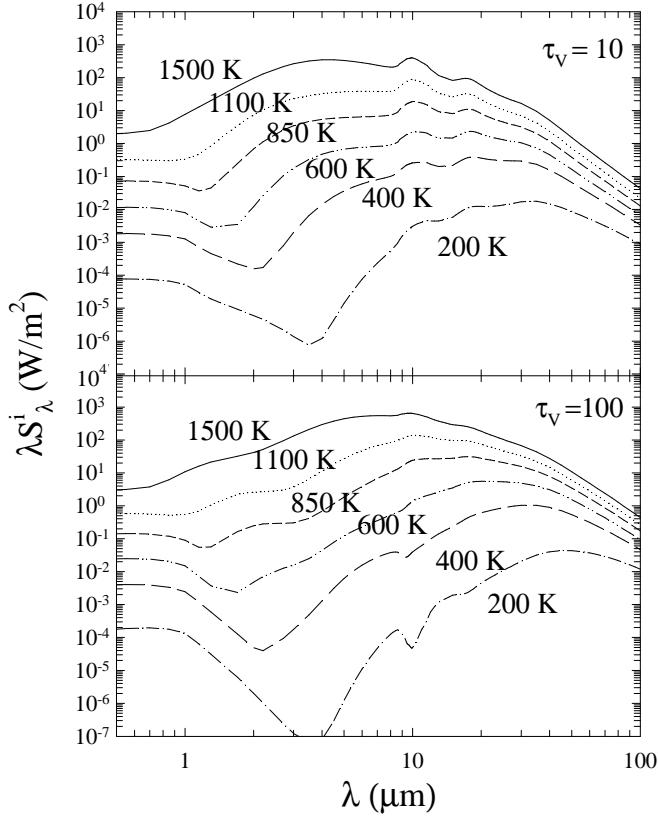}{\figsize}

\caption{Source functions of clouds heated indirectly by the diffuse radiation
for two optical depths and a set of distances, labelled by the equivalent
temperature \Tm.
}
 \label{Fig:Sdif}
\end{figure}


\subsection{Clouds Heated Indirectly}
\label{sec:Indirect}

Clouds whose line-of-sight to the AGN is blocked by another cloud will be
heated only indirectly by the diffuse radiation from all other clouds. Just as
in the standard, smooth density case, the self-consistent solution for the
coupled problems of diffuse radiation field and source function can be readily
obtained by $\Lambda$-iterations. In the first step, calculate the source
functions for all directly heated clouds, and from the emission of these clouds
devise a first approximation for the diffuse radiation field. Next, place
clouds in this radiation field and calculate their emission to derive a first
approximation for the source functions of indirectly illuminated clouds and,
from  eq.\ \ref{eq:S}, the composite source function at every location. In
successive iterations, add to the AGN direct field the cloud radiation
calculated from eq.\ \ref{eq:IC}, and repeat the process until convergence. The
solution must be tested for flux conservation, which takes the same form in
both the smooth and clumpy cases: $\int\! d\lambda \int\! F_{\rm r,\lambda}(r,
\Omega)d\Omega/4\pi = \Fe$, where the angular integration is over a spherical
Gaussian surface of radius $r$ and $F_{\rm r,\lambda}$ is the radial component
of the radiative flux vector, comprised of the AGN transmitted radiation and
the diffuse dust emission. In clumpy distributions, the contribution of the
transmitted flux to this integral becomes \pAGN\Fe, where \pAGN\ is from eq.\
\ref{eq:pAGN}, and the radial component of the diffuse flux at any position is
$F_{\rm r,\lambda}^{\rm C} = \int\IC \cos\theta d\Omega$, where \IC\ is the
intensity from eq.\ \ref{eq:IC} and $\theta$ is angle to the radius vector.
Thus the flux conservation relation becomes
\eq{
    \pAGN\Fe +
    \int_0^1\!\! d\sin\beta \int\! d\lambda\, F_{\rm r,\lambda}^{\rm C}(r,\beta)
            = \Fe\, .
}
With trivial modifications, this is also the result for the smooth-density
case: in the transmitted term (\pAGN\Fe) the clumpy effective optical depth
$t_\lambda$ is replaced by actual optical depth $\tau_\lambda$ in the escape
factor $P_{\lambda, \rm esc}$, and in the diffuse term \IC\ stands for the
smooth-density emission intensity.

\begin{figure}[ht]
 \Figure{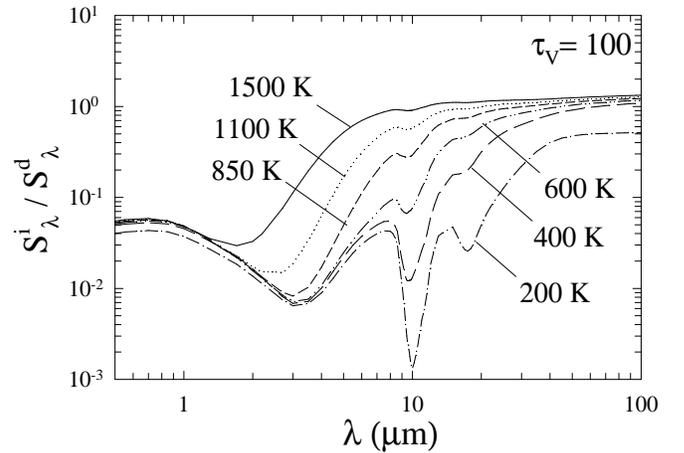}{\figsize}

\caption{Ratio of the source functions for directly- and indirectly-heated
clouds with \tV\ = 100 at the indicated distance-equivalent temperatures.
$S^{\rm d}_\lambda$ is taken at $\alpha = 90\degr$.
}
 \label{Fig:Sfn-ratio}
\end{figure}

Dust is heated predominantly at short wavelengths and is reprocessing radiation
toward longer wavelengths. As a result, at every location the clouds heated
directly by the AGN will be the warmest and provide much stronger heating than
the shadowed, much cooler clouds. The significance of feedback from the clouds
that are heated indirectly can be gauged from $\Si/\Sd$, the ratio of their
source function to the driving term \Sd; when this ratio is small, rapid
convergence can be expected. In our calculations we performed only the first
two steps of the $\Lambda$-iteration process and employed an isotropic
radiation field as an approximation for the heating of shadowed clouds. The
approximate heating field was derived from an average over $\alpha$ of the
emission of directly heated clouds at the given location; this is the radiation
field that would exist inside a spherical shell of directly illuminated clouds
at the given location. In reality, an indirectly heated cloud is probably
exposed to the bright sides of somewhat fewer clouds, because they are on the
far side from the AGN, so our approximation is an upper limit for the strength
of the diffuse field. In this isotropic field we placed dusty slabs and solved
the radiative transfer problem with \D. Figure \ref{Fig:Tdiff} shows the
temperature profile inside such slabs at the location where \Tm\ = 850 K. As
expected, slabs heated by the diffuse radiation are much cooler than their
directly illuminated counterparts, shown in fig.\ \ref{Fig:Td-tau}. As before,
from the radiative transfer solution we obtained \Si\ using the averaging
procedure in eq.\ \ref{eq:clump}. Since now the slab is illuminated
isotropically on both faces, the dependence on \thin\ disappears and \Si\ is
isotropic. Figure \ref{Fig:Sdif} shows results for \Si\ at two representative
optical depths and a range of radial distances, characterized by \Tm. Figure
\ref{Fig:Sfn-ratio} shows the corresponding ratio \Si/\Sd\ for \tV\ = 100. This
ratio is always below \about 10\% around the wavelengths corresponding to peak
emission at each distance; for example, clouds with \Tm\ = 200 are the main
contributors to the emission around \about 15\mic, and as is evident from the
figure, \Si\ is less than 10\% of \Sd\ for this temperature in that spectral
range. Significantly, \Si/\Sd\ is small even though our approximation
overestimates the diffuse radiation strength as it involves an isotropic
distribution of directly heated clouds around every AGN-obscured cloud. As
expected, clouds heated indirectly are much weaker emitters, indicating rapid
convergence of the iteration procedure.


In contrast with figs.\ \ref{Fig:Sdir-tau} and \ref{Fig:Sdir-all}, figure
\ref{Fig:Sdif} shows only a trace of the 10\,\mic\ feature. This disparity
reflects the fundamental differences, evident in figures \ref{Fig:Td-tau} and
\ref{Fig:Tdiff}, between the internal temperature structures of directly- and
indirectly-illuminated clouds. A cloud placed in a black-body radiation field
will thermalize with its temperature $T$ and emit according to $I_\lambda =
B_\lambda(T) (1 - e^{-\tau_\lambda})$. Therefore, when ${\tau_{10\mic} \ll 1}$,
$I_\lambda \simeq B_\lambda \tau_\lambda$ and the emergent spectrum has the
same shape as the dust cross section, producing an emission feature. When
$\tau_{10\mic}$ increases and approaches unity, self-absorption sets in and the
feature strength decreases. As $\tau_{10\mic}$ increases further and exceeds
unity, $I_\lambda$ becomes equal to $B_\lambda$: at constant temperature,
self-absorption and emission exactly balance each other, producing the
thermodynamic limit of the Planck function. For all $\tau_{10\mic}
> 1$, a single-temperature cloud will never produce a feature, either in
emission or absorption. Since our diffuse radiation field differs from a pure
black-body, the temperature inside indirectly heated clouds is not constant,
but its variation is still relatively modest, as is evident from fig.\
\ref{Fig:Tdiff}. While the dust temperature in a directly illuminated slabs
varies by more than factor 4, it varies by less than factor 2 inside slabs
heated from both sides by the isotropic diffuse radiation field. Furthermore,
even this small variation is mostly limited to narrow regions near the heated
surfaces. This explains why fig.\ \ref{Fig:Sdif} shows only a weak emission
feature at \tV\ = 10 (which corresponds to $\tau_{10\mic} \simeq 0.7$) and
practically no feature in most of the \tV\ = 100 clouds.

\section{DISCUSSION}


The formalism presented here is general and can be applied to any clumpy
distribution. In this study we employed it to construct the source functions of
dusty clouds. As a small object, a cloud is primarily characterized by its
overall optical depth \tV. Two additional properties could potentially affect
the cloud emission. One is surface smoothness. A rough, fractal-like surface
can be expected to reduce the efficiency of scattering by absorbing some
scattered photons and transferring them to the thermal pool. The other is
shape. This factor could be studies by averaging ellipsoidal clouds over all
orientations. The shape parameter would then correspond to the ellipsoid axial
ratio as the cloud varied from a spherical shape to extreme elongation; the
clouds constructed here can be considered representative of extremely elongated
shapes. The isotropy of the external radiation field of indirectly heated
clouds enabled us to explore in this case the two extremes of cloud geometrical
shape: since the radiative transfer problem for a spherical cloud retains the
spherical symmetry in an isotropic external field, it too can be solved by \D.
The source function is then found from $\Si = F_\lambda/\Omega$, where
$F_\lambda$ and $\Omega$ are, respectively, the flux and solid angle of the
sphere at a large distance. We calculated \Si\ for spherical clouds imbedded in
the same radiation field used in the calculations described in
\S\ref{sec:Indirect}. The solution for a sphere with uniform density depends
only on the overall optical depth \tV, allowing direct comparison with the
clouds constructed by averaging slabs of the same \tV. We found no significant
differences between the two extreme cases. We are currently studying also the
direct illumination of spherical clouds, and will report our full findings in a
separate publication (Kimball, Ivezi\'c \& Elitzur, in preparation).

Both shape and surface properties appear to be of secondary importance. The
modeling of the source function presented here in which the clouds are
characterized purely by \tV\ does seem to capture the essence of emission from
a single dusty cloud, although definitive conclusions will have to await
detailed comparisons with the calculations for externally illuminated spheres.

\subsection{Dust temperature in clumpy media}

Dust temperature distributions are profoundly different in clumpy and smooth
media. In smooth density distributions, dust temperature and distance from the
AGN are uniquely related to each other---given the distance, the dust
temperature is known, and vice versa. In contrast, as shown in
\S\ref{sec:Tdust}, in a clumpy medium
\begin{itemize}
\item
different dust temperatures coexist at the same distance from the AGN

\item
the same dust temperature occurs at different distances --- the dark side
of a cloud close to the AGN can be as warm as the bright side of a farther
cloud
\end{itemize}
\cite{Schartmann05} find that different components of the multi-grain mix
employed in their torus modeling can have different temperatures at the same
location, therefore the concept of ``dust temperature'' is ill defined at the
microscopic level in such regions. Still, the fundamental differences between
the temperature structures of smooth and clumpy media apply to the individual
components of the grain mix and to the mixture average, as well as to the
common, proper dust temperature when all the components equilibrate to the same
one. These differences have profound implications for the torus emission,
explaining the low dust temperatures found so close to the nuclei of NGC1068
\citep{Jaffe04, Poncelet06} and Circinus \citep{Tristram07}. We discuss these
implications further in part II.

\subsection{X-rays vs IR}

IR flux measurements collect the emission from the entire torus area on the
plane of the sky. This flux is determined by the average number of clouds along
all radial rays through the torus. In contrast, X-ray attenuation is controlled
by the clouds along just one particular ray, the line of sight to the AGN.
Since X-rays are absorbed by dust-free as well as dusty material, X-ray
absorbing clouds will generally outnumber the torus clouds in any given
direction. In fact, X-ray absorption in AGN could be dominated by the dust-free
clouds (see paper II). But even for the dusty portion of the column, the number
of X-ray absorbing clouds can differ substantially from the torus average. As
an example, the appendix table \ref{Table} presents a tabulation of the Poisson
distribution for \N\ = 5, which is representative of AGN tori as shown in part
II. More than 80\% of the paths will have a number of clouds different from 5
in this case, and the probability to encounter just 1 cloud or as many as 9 is
a full 20\% of the probability to encounter the average 5.  Two type 2 sources
with similar cloud properties and the same average \N\ will have an identical
IR appearance, yet the X-ray absorbing columns in each torus could still differ
by an order of magnitude. This can be expected to introduce a large scatter in
X-ray observations.

Each spectral regime responds to large variations in cloud optical depth in an
entirely different way. The IR emission depends on \tV\ through the probability
for photon escape and the cloud source function (see eq.\ \ref{eq:IC}). Both
factors saturate when \tV\ exceeds \about 100. From eq.\ \ref{eq:Pesc}, $\Pesc
= e^{-\cal N}$ when $\tau_\lambda \gg 1$. Since this condition is met at all
relevant wavelengths when $\tV \ga 50$, \Pesc\ becomes independent of \tV.
Similarly for \Sl: Because each cloud is heated from outside, only its surface
is heated significantly when \tV\ is large. Increasing \tV\ further only adds
cool material, thus \Sl\ saturates for all relevant $\lambda$ (similar to
standard black-body emission). Indeed, figure \ref{Fig:Sdir-tau} displays model
results that cover three orders of magnitude of clump optical depth, yet the
SEDs show only moderate variation, saturating altogether when $\tV \ga 100$.
Even at $\tV < 100$, significant spectral variety is mostly limited to $\lambda
\la 10$ \mic\ for clouds along the line of sight to the AGN. In contrast, the
\tV-dependence of X-ray absorption is markedly different. Individual torus
clouds are optically thin to X-rays, because the optical depth for Thomson
scattering is only \about 2\x\E{-3}\tV, therefore the overall optical depth for
X-ray absorption is \N\tV\ (see \S\ref{sec:formalism}). It increases linearly
with \tV, in contrast with the saturated response in the IR regime.

The great disparity between the two spectral regions is expected to further
increase the scatter in torus X-ray properties among AGN with similar IR
emission. It also may help explain why the SEDs show only moderate variations
in the infrared that are not well correlated with the X-ray absorbing columns
\citep[e.g.,][]{Silva04}.

\subsection{The 10\mic\ feature}

In contrast with ultra-luminous infrared galaxies (ULIRGs),  AGN observations
do not provide any example of deep 10\mic\ silicate absorption feature
\citep{Hao07}. This behavior conflicts with the results of smooth-density
models \citep[e.g.,][]{PK92} but is a natural consequence of clumpiness: as is
evident from figures \ref{Fig:Sdir-tau} and \ref{Fig:Sdir-all}, single clumps
never produce extremely deep features. This behavior reflects the flat slab
temperature profile (see fig.\ \ref{Fig:Td-tau}) and is a realistic depiction
of the situation inside an actual cloud heated from outside by a distant
source. As noted by \cite{Levenson07}, the different behavior of the 10\mic\
absorption in ULIRGs and AGN indicates that the dust distribution is smooth in
the former and clumpy in the latter \citep[see also][]{Spoon07}.

Clumpiness suffices by itself to explain the modest depth of the 10 \mic\
absorption feature in AGN. The complete behavior of the feature, including the
transition from emission to absorption, involves an intricate interplay between
the relative contributions of clouds at different locations and their mutual
shadowing. This behavior displays a complex pattern that depends on the actual
geometry of the cloud distribution. A detailed, quantitative analysis of the
10\mic\ feature in clumpy tori is performed in part II.

\subsection{Conclusions}

Clumpy media differ from continuous ones in a number of fundamental ways. The
low dust temperatures found close to the nucleus of NGC1068 contradict basic
physical principles in smooth density distributions but arise naturally in
clumpy ones. Two additional puzzling features of IR emission from AGN are
simply hallmarks of clumpy dust distributions, independent of the distribution
geometry: Even if the cloud configurations were spherical or irregular rather
than toroidal, (1) the SED still would show only a moderate range
incommensurate with the variation of X-ray attenuation, and (2) the 10\mic\
absorption feature would never be deep. Understanding the full range of 10\mic\
features observed in AGN spectra requires considerations of the actual geometry
of the cloud distribution, though. This problem is addressed in part II
together with other implications of clumpy tori to AGN observations.

\acknowledgements
Part of this work was performed while M.E.\ spent a most enjoyable sabbatical
at LAOG, Grenoble. We thank Nancy Levenson, Paulina Lira and Hagai Netzer for
their useful comments on the manuscript. Partial support by NSF and NASA is
gratefully acknowledged.

\appendix
\section{Poisson Statistics}

\begin{table}[ht]
\begin{center}
\caption{}
\begin{tabular}{rccc}
  \hline
 $k$   &  $P_k$  & $P_k/P_{\cal N}$ & $\sum\limits_{i \le k}P_i$ \\
 \hline
 0   & 0.0067 & 0.04 & -- \\
 1   & 0.0337 & 0.19 & 0.04 \\
 2   & 0.0842 & 0.48 & 0.12 \\
 3   & 0.1404 & 0.80 & 0.27 \\
 4   & 0.1755 & 1.00 & 0.44 \\
 5   & 0.1755 & 1.00 & 0.62 \\
 6   & 0.1462 & 0.83 & 0.76 \\
 7   & 0.1044 & 0.60 & 0.87 \\
 8   & 0.0653 & 0.37 & 0.93 \\
 9   & 0.0363 & 0.21 & 0.97 \\
 10  & 0.0181 & 0.10 & 0.99 \\
 11  & 0.0082 & 0.05 & 0.99 \\
 12  & 0.0034 & 0.02 & 1.00 \\
 \hline
\end{tabular}
\end{center}
\tablecomments{Poisson probability $P_k$ (second column) for the number $k$ listed in the
first column when the average is \N\ = 5. The third column normalizes these
probabilities to the one for hitting the average \N. The last column lists the
cumulative probability.}
\label{Table}
\end{table}

The elementary problem most relevant for the statistics of clouds along a ray
is the distribution of points placed randomly on a circular board. Denote by
$n$ the total number of points and by $p\ (\ll 1)$ the probability that a point
lands on a given radial ray. Then the probability for $k$ points to land on
that ray is
\eq{
    P_k = {n!\over k!(n - k)!}\,p^{k}(1 - p)^{n - k}
}
If \N\ is the average number of points landing on a ray than $p~=~\N/n$,
therefore
\eq{
    P_k = {\N^k\over k!}\left(1 - {\N\over n}\right)^n
    \times\left[\left(1 - {1\over n}\right)\!\!\left(1 - {2\over n}\right)
          \ldots\left(1 - {k + 1\over n}\right)
          \left(1 - {\N\over n}\right)^{-k}\right]
}
In the limit $n \to \infty$ with \N\ and $k$ fixed, every term in the square
brackets approaches unity while $(1 - \N/n)^n \to e^{-\cal N}$, yielding the
Poisson distribution
\eq{\label{eq:Poisson}
    P_k \simeq {\N^k\over k!} e^{-\cal N}.
}
Equation \ref{eq:Pesc} for the photon escape probability follows immediately.
The probability to encounter $k$ clouds along the path is $P_k$. If the optical
depth of each cloud is $\tau$ then the probability to escape from all of these
encounterers is $e^{-k\tau}$. Therefore
\eq{
  \Pesc = \sum_k P_k e^{-k\tau}
  = e^{-\cal N}\sum_k {(\N e^{-\tau})^k\over k!}
}
yielding the Natta \& Panagia result (eq.\ \ref{eq:Pesc}).

It is important to note that the only requirement for eq.\ \ref{eq:Poisson} is
that the total number of points obey $n \gg k, \N$; the mean number \N\ of
points on a ray can be small (even less than unity). While the statistics of
points along a ray always follows the Poisson distribution, there is no similar
universal limit for the statistics of the average \N, and it remains arbitrary.
As an example, consider a large number of identical boards with the same number
of points $n$ spread on each one of them, so that the average \N\ is the same
for every board. In this case \N\ will have a $\delta$-function distribution
while the number of points on any given ray in each board will obey the Poisson
distribution around that common average.

Finally, the Poisson distribution allows significant deviations from the
average. From the accompanying tabulation for \N\ = 5, the probability to hit
just one cloud or as many as nine clouds is 20\% of the probability to hit the
average five clouds. This implies a substantial probability that the number of
clouds along any particular line of sight, such as the one to the AGN, will
deviate significantly from the torus average \N.




\end{document}